\documentclass[aps]{revtex4}
\begin{document}

\title{THE BOSE GAS LOW MOMENTUM LIMIT REVISITED}
\author{H. Perez Rojas$^{1,2}$and D. Oliva $^{2}$}
\address{\vskip 1.5cm $^1$International Center for Theoretical Physics,\\
P.O. Box 586 34100 Trieste, Italy\\
$^2$ Grupo de Fisica Teorica, ICIMAF,\\
Calle E No. 309, Vedado, La Habana, Cuba.}

\begin{abstract}
We discuss the standard approach to the problem of the low
momentum limit of the spectrum for a weakly interacting Bose gas.
The Bogoliubov´s spectrum is shown to be obtained as a Goldstone
mode thanks to the introduction of a chemical potential $\mu$.
This procedure has, however, difficulties since the breaking of
the gauge symmetry implies that the corresponding chemical
potential must be taken as zero, unless it is introduced before
breaking the symmetry. But if this is done, after the symmetry
breaking $\mu$ loses its meaning as a chemical potential. An
alternative two-mode solution is suggested having two modes, one
of them being the free-particle quadratic in momentum spectrum,
the second bearing a gap. This gap leads to a $\lambda$-type
behavior of the specific heat near the critical temperature.
\end{abstract}
\maketitle

\vskip 0.5cm

\section{Introduction}
Most studies for the weakly interacting Bose gas deal with the zero
temperature approximation. We must cite at first the famous Bogoliubov model
\cite{Bogoliubov}. In it the assumption $N - n_0 = \sum_{{\bf p}} n_{{\bf p}%
} \ll N $ is made, so that one can discard a term of order $(\sum_p n_{{\bf p%
}})^2$. This leads to a gapless spectrum, but such approximation
is actually valid for temperatures extremely close to $0
^{\circ}K$. Bogoliubov's sprectrum is in agreement with the pure
phenomenological predictions made by Landau concerning the
superfluid spectrum, \cite{Landau} whose lower branch is expected
to have a phonon behavior. Far from $0^{\circ}K$, as happens near
and below the critical temperature $T_c$, the approximation made
by Bogoliubov is, however, not valid, and the spectrum bears a gap
\cite{David}, since it is,
\begin{equation}
\epsilon(p)= \sqrt{(p^2/2m)^2 + 8K(p^2/2m) + 12K^2},  \label{sp}
\end{equation}
\noindent
where $K =U_0 n_0/2V$, and $U_0$ is the (approximately) constant potential,
$n_0$ the condensate and $V$ the volume. As shown in \cite{David},
this gap, not being satisfactory with the expected spectrum, bears
however the interesting property of leading to a divergent
behavior of the heat capacity of the system, which fact is in correspondence with
the well-known lambda-type phase transition of He$^4$. The usual Bogoliubov
gapless spectrum does not  predict any $\lambda$ behavior \cite{Huang} and is
valid only near $T=0$, also it does not exhibit even the discontinuity in the $C_v$
derivative at $T_c$ typical of the usual ideal Bose gas.

 Bogoliubov's work was a brilliant insight in the
appearance of gapless particles after spontaneous breaking of a
gauge symmetry, a fact which became more clear 15 years later,
after the Goldstone theorem \cite{Goldstone}. After Bogoliubov the
weakly interacting gas spectrum was investigated by Beliaev
\cite{Beliaev}, by using the Green's function method, but the
correspondence with Bogoliubov's results could be obtained only
after introducing a quantity identified as a chemical potential,
which was not taken into account in the original model by
Bogoliubov. Below we shall see, by a different method than the one
used by Beliaev, that spontaneous symmetry breaking is produced by
introducing explicitly the chemical potential, leading to the
appearance of a gapless Goldstone mode. On the other hand, by
taking the chemical potential as equal to zero in Beliaev's
calculations, the spectrum (\ref{sp}) is reproduced.

In the present letter we are interested essentially in discussing
the meaning of taking such a chemical potential as different from
zero, at the light of the fundamental principles of statistical
mechanics.
\section{The hamiltonian}
Condensation breaks the gauge symmetry (see below). The
introduction of a chemical potential coupled to a Noether charge
$Q$ associated to a broken symmetry in the density matrix was
investigated in the relativistic context in papers
\cite{chaichian}. Such charge has many odd features from which we
mention two: first, in general, it is not conserved, and second,
its introduction ban the fulfillment of the Kubo-Martin-Schwinger
(KMS) condition required for the statistical description of
systems in thermodynamical equilibrium, since according to the
Araki-Haag-Kastler-Takesaki (AHKT) theorem \cite{Haag}, only those
charges that belong to the center of the unbroken part of the
symmetry group satisfy the KMS condition and can have a chemical
potential associated to them.

In the non-relativistic weakly interacting Bose gas case, the
charge operator associated to the broken symmetry is $N=\int
d^3x\psi ^{*}(x)\psi (x)=\sum_{p=0}^{p\to \infty }a^{*}(p)a(p)$,
where it is taken the condensate density as $a^{*}(0)a(0)=n_0$ and
in consequence $a(0)=\sqrt{n_o}e^{i\theta }
$,$a^{*}(0)=\sqrt{n_o}e^{-i\theta }$.This means \cite{strocchi}
that the physically relevant representations of the algebra of
canonical variables are labelled by the parameters $n_0,\theta $,
and
\begin{equation}
lim_{V\to \infty }\int d^3x<n_0,\theta |\psi (x)|n_0,\theta
>=\sqrt{n_0}e^{i\theta }
\end{equation}
which for $n_0\neq 0$ is not invariant under gauge transformations

Let us consider the Lagrangian
\begin{equation}
{\cal L}= -\psi^{*}(x)i \partial_0\psi(x) + \psi^{*}(x) \frac{\nabla^2}{2m}%
\psi(x) - \lambda(\psi^{*}(x)\psi(x))^2
\label{lagrangian}
\end{equation}
\noindent where $\lambda = \frac{U_0}{2V}$ and we are considering
the approximation of a constant repulsive potential. Under the
gauge transformation
\begin{equation}
\psi (x)\to e^{i\alpha }\psi (x)
\end{equation}
terms having equal number of $\psi (x)$ and $\psi ^{*}(x)$ are
gauge invariant, so ${\cal L}$ is gauge invariant, as it is also
the density of particles $N$.

From the Lagrangian we can get immediately the Hamiltonian in
which, by taking the Schrodinger representation of the fields
$\psi (x)=\frac{1}{V}\sum a(p)e^{i \textbf{p}\cdot
\textbf{r}/\hbar}$, $\psi^{*} (x)=\frac{1}{V}\sum a^{*}(p)e^{-i
\textbf{p}\cdot \textbf{r}/\hbar}$, one obtains in momentum space
an expression quadratic in the fields $a^{*}(p)$ , $a(p)$ . After
condensation, if it is taken  $a^{*}(p)=a^{*}(p)e^{-i\theta}$ and
$a(p)= a(p)e^{i \theta}$ obviously, such a Hamiltonian is not
gauge invariant. If we fix the gauge $\theta = 0$, we get finally
the bilinear part of the Bogoliubov's Hamiltonian,
\begin{eqnarray}
H_B& = & \left. \lambda n_0^2 + \sum_{p}[a^{*}(p) \frac{p^2}{2m}a(p) + K(4
a^{*}(p)a(p) + a^{*}(p)a^{*}(-p) + a(p)a(-p)]\right.
\end{eqnarray}
\noindent
where $K= \lambda n_0$. Note that the coefficient of the first "potential"
term is $4K$ and {\it not} $2K$, as is done as an approximation in the usual
treatment of the problem in terms of the total number of particles $N$.
\section{Density Matrix and Beliaev equations}
The immediate task is to build a density matrix aimed to obtain
the partition functional ${\cal Z}$. If we name $N_p = \sum_p
a^{*}(p)a(p)$ the operator for the number of excited particles, we
see that $[H_B, N_p] \neq 0$. In consequence, there is no a common
set of eigenstates for these two operators and the density matrix
$\rho = e^{-(H_B - \mu N_p)\beta}$ is an ill-defined one for
describing states of equilibrium. If we define
$\Omega=-\beta^{-1}\ln {\cal Z}$ one cannot state that
$N=-\frac{\partial \Omega_{\beta}}{\partial \mu}$, is the average
number of excited particles.
However, one might argue that we have the right to consider $\mu
\neq 0$ as a Lagrange multiplier, since in building $\rho$ and
$\Omega$ it loss the meaning of a chemical potential but then we
face new difficulties.

Working according to Bogoliubov's standard procedure
\cite{Landau}, we may proceed to diagonalize $H_B$,
\begin{equation}
H_B = \sum_{{\bf p}}\epsilon(p)b^{*}(p)b(p)
\end{equation}
where according to Bogoliubov's transformation, we have $b^{*}(p) = [a(p) +
\alpha_p a^{*}(-p)]/(1 - \alpha^2 (p))$, $b (p) = [a^{*}(p) + \alpha_p a
(-p)]/(1 - \alpha^2 (p))$, and $\alpha (p) = [4K +p^2/2m-\epsilon (p)]/2K$).
In this representation, in which $b^{*}(p)b(p)$ is gauge invariant under the
transformation $b(p) \to e^{i \theta} b(p) $, $b^{*}(p) \to e^{-i \theta}
b^{*}(p)$, we have
\begin{equation}
N = \sum_p \frac{(1 + \alpha (p)^2)n_p + \alpha (p)^2}{1 - \alpha (p)^2} -%
\frac{\alpha_p}{1 - \alpha (p)^2}(b^{*}(p) b^{*}(-p) + b(p) b(-p)),
\end{equation}
\noindent
which is obviously not gauge invariant. The conditions of the AHKT theorem
\cite{Haag} are not satisfied. Thus, the system of equations (\ref{minxi})
and (\ref{numpart}) are also not gauge invariant.

Thus if we continue in a pure formal way, and leave $\mu \neq 0$, by
considering $H_B - \mu (N_p+n_0)$ as a new Hamiltonian, one can obtain some partition function $%
{\cal Z}$ from it, from which the system of Green functions and
the thermodynamical  potential $\Omega$ can be obtained. By
introducing adequate external sources, we have for the partition
function obtained from the unperturbed Hamiltonian $H_B$,

\begin{equation}
{\cal Z}(\eta, \zeta) = M(\beta)e^{-\beta V \lambda n_0^2-\mu n_0}{\cal %
Z_{\beta}}
\end{equation}
\noindent where $M(\beta)$ is a temperature dependent constant and
\begin{equation}
{\cal Z_{\beta}}= \int{\cal D}a^{*}{\cal D}a e^{\sum_{p_4, p} [S(p) +\eta
(p) a^{*}(p)+ \zeta(p)a(p)]}  \label{part}
\end{equation}
\noindent
and
\begin{eqnarray}
S(p)& = &\left.(i p_4 + \mu - \frac{p^2}{2m})a^{*}(p)a(p) - K(4 a^{*}(p)a(p)
+a^{*}(p)a^{*}(-p) + a(p)a(-p))\right.  \label{ac}
\end{eqnarray}
We may consider the interaction term in the Hamiltonian,
containing terms of third and fourth order in $a^{*}(p), a(p)$. If
$V(a(0), a^{*}(0), a^{*}(p), a(p) )$ is such term, the interacting
partition function is ${\cal Z}_I=e^{\sum_{p_4, p}V(a(0),
a^{*}(0),\frac{\delta}{\delta \eta (p)}\frac{\delta}{\delta \zeta
(p)})}{\cal Z}$. By using the standard methods of functional
differentiation, we may develop a formalism  to obtain from ${\cal
Z}_I$ the equations for the Green functions $G(p,p^{\prime
})=\delta ^2{\cal Z}_I/\delta \zeta (p)\delta \eta (p^{\prime })$, $%
G_1(p,p^{\prime })=\delta ^2{\cal Z}_I/\delta \eta (-p)\delta \eta
(p^{\prime })$. We will restrict, however, to the corresponding
expressions for the $V=0$ approximation ${\cal Z}_I={\cal Z}$, by
taking
 the bilinear terms in (\ref{ac}), and by calling $%
G_0^{-1}=(ip_4+\mu -\frac{p^2}{2m})$ one easily gets,
\begin{equation}
G(p)=G_0(p)+4KG_0(p)G(p)+2KG_0(p)G_1(p)
\end{equation}

\begin{equation}
G_1(p) = 2K G_0 (-p) G(p)+ 4K G_0 (-p) G_1 (p)
\end{equation}
These are a simplified version of the Beliaev's equations \cite{Beliaev},
which lead to
\begin{equation}
G_1(p) = \frac{2K G (p)}{G_0^{-1} (-p) - 4K}
\end{equation}
\begin{equation}
G(p) = \frac{G_0(-p) - 4K}{(p_4 - i\epsilon (p))(p_4 + i \epsilon (p))}
\end{equation}
\noindent
where
\begin{equation}
\epsilon(p) = \sqrt{(\frac{p^2}{2m}-\mu + 4K)^2 -4 K^2}.  \label{esbeli}
\end{equation}
The effective potential is obtained as $U = \beta^{-1}\ln {\cal
Z}/V$, by defining $\xi = \sqrt{n_0}$, as
\begin{equation}
U = \lambda \xi^4 -\mu \xi^2 + \Omega_{\beta}  \label{efpot}
\end{equation}
\noindent where $\Omega_{\beta} =\ln {\cal Z_{\beta}}$. We see
that the chemical potential $\mu$ plays an essential role in
showing explicitly the symmetry breaking term in $U$, analogous to
the "negative mass term" of the usual relativistic theory
\cite{Goldstone}. Its presence leads necessarily to the existence
of Goldstone bosons.

From (\ref{efpot}) one can obtain the condition of extremum of
$U$,
\begin{equation}
\frac{\partial U}{\partial \xi}= 4\lambda \xi^3 -2\mu \xi
+\frac{\partial \Omega_{\beta}}{\partial \xi} = 0,  \label{minxi}
\end{equation}
\noindent But we cannot write that the quantity
\begin{equation}
-\frac{\partial U}{\partial \mu} = n_0 + \frac{\partial
\Omega_{\beta}}{\partial \mu}  \label{numpart}
\end{equation}
\noindent as the average number of particles, i.e,
$N_q=-\frac{\partial \Omega_{\beta}}{\partial \mu}$ is not the
number of excited particles. Usually it is interpreted as the
average number of "quasiparticles". For $T \to 0$,$N_q \to 0$ and
only in that case $n_0 \sim N$.

The equation (\ref{minxi}) gives $\xi^2 = \mu/2 \lambda$ as the
minimum at the tree level, i.e., for $T = 0$ $\mu= 2K$. By
substituting in
(\ref{esbeli}), this leads immediately to the (gapless) Goldstone boson $%
\epsilon(p)_B = \sqrt{(\frac{p^2}{2m})^2 +4 K \frac{p^2}{2m}}$,
which is the usual Bogoliubov spectrum. According to Goldstone
theorem, these modes if calculated correctly, would maintain
gapless at all orders of perturbation theory. This result is in
correspondence to the one obtained, before the formulation of the
Goldstone theorem, by Hugenholtz and Pines \cite{Pines}.

We see that for the reasons pointed out above, there is no right
to
build a density matrix in the grand canonical ensemble by subtracting from $%
H_B$ the operator $N_p$ multiplied by a nonzero chemical potential
$\mu$, since as $[H_B, N_p] \neq 0$, there is no a common set of
eigenstates and eigenvalues for them, and $\rho = e^{-(H_B - \mu
N_p)\beta}$ does not satisfy the KMS condition. Thus, it is
unsatisfactory to attribute a standard physical meaning to quantities derived
from it if $\mu \neq 0$, in particular, to
interpret $\frac{\partial \Omega_{\beta}}{\partial \mu}$ in (\ref{numpart}%
) as the average number of excited particles. We introduce the
notion of "quasiparticles" in part to avoid such a difficulty. We
point out also that if we take $\mu = 0$ (\ref{esbeli}) turns into
(1).
\section{Alternative procedure}

We conclude that the standard introduction of a chemical potential
is not satisfactory if it is done after breaking the gauge
symmetry. If we include it before the symmetry breaking, it loss
its meaning as a chemical potential and becomes a parameter
essential for the fulfillment of the Goldstone theorem. However it
leads (according to the gauge fixing) to other infinite
non-equivalent set of representations of the operator algebra, and
of their dynamics \cite{strocchi}. There is no  a priori criteria
to state which of these representations is physically better. One
may choose, for instance, that one wants the energy eigenvalue $
E(p)$ to be linear in $p$ for $p\longrightarrow 0$. Then one
possible solution is the Bogoliubov-Beliaev spectrum. But from the
point of view of Goldstone theorem and its consequences, the use
of the fields $\psi ^{*}(x)$ and $\psi (x)$, both containing the
symmetry-breaking parameter, is rather obscure. Only one physical
mode is present. It would ban the appearance of a pure gapless
mode and a mode bearing a gap, which is to be expected (fron the
Goldstone theorem point of view) since we
are using two-component fields.

If we do not tie to the linear mode requirement, an alternative
model is, for instance, the one obtained by taking in
(\ref{lagrangian}) $\psi =\xi +\sigma (x)+i\ h(x)$. At the tree
level these two modes have the spectra $E_{h}=p^{2}/m$ and
$E_{\sigma }=p^{2}/m+\Delta _{\sigma }$ where $\Delta _{\sigma} =
K=\lambda n_0$ and $n_0$ is the condensate (which is the total
number of particles at $T=0$) The first mode has the property of
being gapless, and trivially corresponds to the Goldstone Boson
\cite{Goldstone}, since $E_h \longrightarrow 0$ for
$p\longrightarrow 0$.  The second mode bears a gap. At $T\neq 0$,
this gap is expected to be of form $\Delta _{\sigma }(T)= \lambda
\ n_{0}\sqrt{1-(\frac{T}{T_{c}})^{3}}$, where $T_c$ is the
critical temperature for condensation. It bears the interesting
property of leading to a $\lambda$ - behavior near $T_c$,as shown
below. We observe that the Bogoliubov-Beliaev spectrum is just the
geometric average of the spectra $E(p)=\sqrt{E_{\sigma } E_{h}}$ this
spectrum does not lead to an $\lambda$-type behavior.

The divergence of the specific heat $c_V$ is seen \cite{David}
easily starting from $E_\sigma$. One can write the thermodynamic potential as $\Omega
=AT\int_0^\infty p^2dp\ln (1-e^{-\epsilon_n \beta })$, where $A$
is a constant. By taking $x=p/\sqrt{ 2 m T}$, in the infrared
limit, where $p_0$ is a sufficiently small momentum, we have
\begin{equation}
\Omega \sim -\frac A3 \int_0^{p_0} \frac{p^4 dp}{m (e^{\epsilon_n
\beta }-1)} \sim -\frac{ A(2mT)^{5/2}}{3m}\int_0^{p_0}
\frac{x^4dx}{x^2+\Delta_{\sigma}T^{-1}}\sim
-\frac{A(2m)^{3/2}T\Delta_{\sigma)}^{3/2}\pi }6.
\end{equation}
\noindent From $c_V=-T\frac{\partial ^2\Omega }{\partial T^2},$ we
get that $c_V$ diverges as $\Delta_{\sigma}^{-1/2}$ for $T \to
T_c$.

It may be argued however, that the existence of a gap, as well as
the non-linear gapless mode, contradicts the generally accepted
idea of a phenomenological linear in $p$ phonon spectrum
\cite{Landau}. One concludes that the model we are considering is
an oversimplification of the real system, an that one must study
the higher order corrections. Work in this sense is in progress.
Another modification would be to introduce some term accounting
for the oscillatory motion of the condensed or quasi-condensed
atoms (the spectra obtained above refers actually to translational
modes), leading to a phonon field having linear behavior of the
spectrum. New models for superfluidity are currently being
elaborated \cite{Zheng} on much more phenomenological basis by
including such vibrational terms.

\section*{Acknowledgements}

The work have been supported by \emph{Ministerio de Ciencia
Tecnolog\'{\i}a y Medio Ambiente} under the grant CB0407.  The
authors would like to thanks the invitation and  hospitality of
the Abdus Salam International Center for Theoretical Physics ICTP
and in particular to  Prof V. Kravtsov, Head of the Condensed
Matter Section, allowing the visit to the Center of one of the
author (D.O.A).


\begin{references}
\bibitem{work in progress}work in this problem is in progress by
         the present authors.
\end{references}
\end{document}